\documentclass[fleqn,usenatbib]{mnras}

\usepackage{newtxtext,newtxmath}

\usepackage[T1]{fontenc}
\usepackage{ae,aecompl}
\usepackage[dvipsnames]{color}

\usepackage{graphicx}	
\usepackage{amsmath}	
\usepackage{amssymb}	

\title[Millimetron and EHT VLBI]{Simulations of M87 and Sgr A* Imaging with the Millimetron Space Observatory on near-Earth orbits.}

\author[A. S. Andrianov et al.]
{A.~S.~Andrianov$^{1}$, A.~M.~Baryshev$^{2}$, H.~Falcke$^{3}$, I.~A.~Girin$^{1}$, T. de Graauw$^{1}$,
\newauthor V.~I.~Kostenko$^{1}$, V.~Kudriashov$^{3}$, V.~A.~Ladygin$^{1}$, S.~F.~Likhachev$^{1}$, F.~Roelofs$^{3}$,
\newauthor A.~G.~Rudnitskiy$^{1}$\thanks{E-mail: arud@asc.rssi.ru}, A.~R.~Shaykhutdinov$^{1}$, Y.~A.~Shchekinov$^{1}$, M.~A.~Shchurov$^{1}$
\\
$^{1}$Astro Space Center, Lebedev Physical Institute, Russian Academy of Sciences, Profsoyuznaya str. 84/32, Moscow, 117997, Russia\\
$^{2}$Kapteyn Astronomical Institute, University of Groningen, P.O. Box 800, 9700 AV Groningen, the Netherlands\\ 
$^{3}$Research Institute for Mathematics, Astrophysics and Particle Physics (IMAPP), Radboud University, Heyendaalseweg 135,\\6525 AJ Nijmegen, the Netherlands\\}

\date{Accepted XXX. Received YYY; in original form ZZZ}

\pubyear{2020}

\begin{document}
\label{firstpage}
\pagerange{\pageref{firstpage}--\pageref{lastpage}}
\maketitle

\begin{abstract}
High resolution imaging of supermassive black holes shadows is a direct way to verify the theory of general relativity at extreme gravity conditions. Very Long Baseline Interferometry (VLBI) observations at millimeter/sub-millimeter wavelengths can provide such angular resolution for supermassive black holes, located in Sgr A* and M87. Recent VLBI observations of M87 with the Event Horizon Telescope (EHT) has shown such capabilities. The maximum obtainable spatial resolution of EHT is limited by Earth diameter and atmospheric phase variations. In order to improve the image resolution longer baselines are required. Radioastron space mission has successfully demonstrated the capabilities of Space-Earth VLBI with baselines much larger than Earth diameter. Millimetron is a next space mission of the Russian Space Agency that will operate at millimeter wavelengths. Nominal orbit of the observatory will be located around Lagrangian L2 point of the Sun-Earth system. In order to optimize the VLBI mode, we consider a possible second stage of the mission that could use near-Earth high elliptical orbit (HEO).
In this contribution a set of near-Earth orbits is used for the synthetic space-ground VLBI observations of Sgr~A* and M87 in joint Millimetron and EHT configuration. General-relativistic magnetohydrodynamic models (GRMHD) for black hole environment of Sgr~A* and M87 are used for static and dynamic imaging simulations at 230 GHz. A comparison preformed between ground and space-ground baselines demonstrates that joint observations with Millimetron and EHT significantly improve the image resolution and allow the EHT+Millimetron to obtain snapshot images of Sgr~A* probing dynamics at fast timescales.
\end{abstract}



\section{Introduction}
Light from a luminous accretion disk around black hole propagating at a distance within a few Schwarzschild radii forms at the image plane a shadow with the size of $\sim 5r_{S}$ \citep{Bardeen1973a, Bardeen1973b, Luminet1979}. The precise shape of the shadow depends on the black hole space-time geometry around the black hole and the distribution of emission of the accretion disk.

Currently, the best candidates for the study of the space-time geometry around the black hole are Sgr A* and M87. With a mass of $\sim 4.3 \times 10^6M_{\odot}$ and at a distance of $\sim 8.3$ kpc \citep{Ghez2008, Gillessen2009, 2018yCat..36189010G, 2019Sci...365..664D}, the expected apparent size of the shadow of Sgr~A* is about 53 $\mu$as. Supermassive black hole in M87 has a larger mass of $\sim 3.5-6.6\times 10^9M_{\odot}$ \citep{Gebhardt2011, Walsh2013, EHT2019_p6}, but located further away from Earth at a distance of $\sim$ 17~Mpc \citep{Bird2010}. Consequently, the apparent size of the shadow of M87 is similar to that of Sgr~A* and is about $\sim 40$~$\mu$as.

Far-infrared (FIR) and sub-millimeter wavebands are optimal for imaging the shadow, because ambient plasma is still transparent and refractive scattering decreases at these wavelengths \citep{Falcke2000, Moscibrodzka2009, Moscibrodzka2014, Moscibrodzka2016, Dexter2012}. 

Imaging the shadow has a great importance for testing the theory of general relativity \citep[see for more recent discussion in ][and references therein]{Psaltis2018, Berti2019, Cunha2018}. As has been shown in \citep{bromley2001, broderick2009, Moscibrodzka2009, dexter2010,Dexter2012, kamruddin2013, Broderick2014}, the most indicative feature of the shadow vicinity of a SMBH is an asymmetric photon ring, whose asymmetry encodes the most important parameters of the space-time metric around: the SMBH mass, the spin and its inclination relative to the observer's line of sight. The presence of a crescent-shaped structure around the horizon area in M87 has been robustly confirmed by EHT in their 2017 campaign, with a pronounced asymmetry -- clearly varied thickness along the bright ring, and a deep (more than factor 10) brightness depression towards the center \citep{2019ApJ...875L...1E,2019ApJ...875L...5E,EHT2019_p6}.

However, the main characteristic of the asymmetry -- the mean difference between the outer and inner radius of the ring, is too small -- order of the gravitational radius $\sim 4~\mu $as, and could not be resolved by the EHT \citep[see in ][]{EHT2019_p6}. As discussed recently in \citep{Johnson2019} interferometry with longer baselines -- 2 to 3 orders of magnitude of the Earth size, are required for the detailed study of space-time geometry and radiation transfer at the edge having the same observational frequency of 230 GHz. In addition to the limited EHT resolution, the effects of direct emission from the surrounding plasma make it difficult to accurately measure the photon ring with the current EHT array \citep{2019ApJ...875L...4E, 2019ApJ...875L...5E, Johnson2019}.

Firm determination of critical parameters of the black holes in Sgr~A* and M87 will require much higher angular resolution than can be obtained with ground based interferometry. This can be achieved with either Space-Space VLBI as described recently by \citet{Roelofs2019}, or Space-Earth VLBI that is possible within a joint program between EHT and Millimetron Space Observatory (MSO). In this paper we describe synthetic observations of Sgr~A and M87 with the joint EHT-MSO Space-Earth VLBI. The paper is organized as follows. In the next two Sections \ref{eht} and \ref{mso} we briefly review the EHT collaboration and Millimetron space observatory. Section \ref{siml} describes details of our simulations, including source models, MSO capabilities and orbit configurations. Description of synthetic observations both averaged and dynamic are given in Section \ref{res}, while Section \ref{summ} summarizes the results.

\subsection{Event Horizon Telescope}\label{eht}

The most recent effort to image black hole shadow was done by the Event Horizon Telescope (EHT) collaboration which is a global ground millimeter VLBI array. The primary goal of the EHT is to observe the close environment of the supermassive black hole Sagittarius A* (Sgr~A*) at the center of our Galaxy and the black hole in the center of the giant elliptical galaxy M87 \citep{Doeleman2009, Fish2013, Goddi2017}.  

Previous EHT measurements have constrained the size of Sgr~A* and M87 at 1.3 mm, but did not have sufficient $(u, v)$ coverage for reconstructing an image of the source \citep{Doeleman2008, Doeleman2012, Johnson2015, Lu2018}. Phase closure measurements have indicated asymmetry in the structure of Sgr~A* \citep{Fish2016, Lu2018}. 

The first EHT imaging observations were conducted in April 2017. These observations gave the first horizon-scale resolved image of M87. The mass of black hole was measured and the spin was constrained to be $> 0$ by requiring GRMHD simulations in the EHT model space that reproduce the 1.3 mm structure to also produce a jet with enough power as observed on large scales \citep{2019ApJ...875L...1E, 2019ApJ...875L...2E, 2019ApJ...875L...3E, EHT2019_p6, 2019ApJ...875L...5E}. The EHT has opened a new era in physics and astronomy by enabling the direct study of spacetime and physical processes at the edge of the event horizon of black hole. At the same time a rather restricted angular resolution ($\approx$25 $\mu$as) reachable on ground-based VLBI does not look sufficient for a cogent solution of precise measurements of asymmetry, mass and spin of black hole. This circumstance urges new observational technologies with the Space-Earth VLBI (S-E VLBI).

\subsection{Millimetron Mission}\label{mso}

The most recent space-Earth VLBI has been implemented in Radioastron space mission, which is a 10 meter space radio telescope, that operated successfully for more than 7 years \citep{Ra_2013, Ra_2015, Ra_2017}. Radioastron formed a space-Earth interferometer together with up to 60 ground telescopes at four frequencies: 0.3~GHz, 1.6~GHz, 4.8~GHz and 22~GHz. It used elliptical orbit with minimal perigee of 400 km and maximum apogee $\approx$ 330000 km \citep{RA_Orbit}.

Observations with the Radioastron showed, that relatively compact AGNs are seen at 22~GHz with flux density sufficient for the detection even at baselines, that correspond to angular resolution of 11 $\mu$as and the best sensitivity achieved at baselines Radioastron-Green Bank telescope at 1 cm was $\approx 10$ mJy \cite{Kovalev2020}.

Millimetron observatory will be a deployable 10 meter cooled far infrared, sub-millimeter and millimeter space telescope \citep{MM_2014}. During the launch the primary mirror and cryogenic screens will be folded in order to fit under the launcher fairing. In contrast to Radioastron, Millimetron observatory will be operating in two modes: single dish and space-ground interferometer. 

In the single-dish mode, Millimetron will measure CMB spectral distortions, magnetic fields, observe filamentary structure and the water trail in the Galaxy.

As a part of Space-Earth interferometer, the goal of Millimetron is to provide high angular resolution for millimeter VLBI, that is crucial for the studies of very compact astrophysical objects like black holes. Space-VLBI mode will be used to observe in the wide frequency range from 33 up to 720 GHz (see Table~\ref{Table_MM}).

The expected sensitivity of Millimetron (due to 125 times wider bandwidth and higher effective antenna area) will be orders of magnitude greater than the sensitivity of Radioastron. These facts allow us to expect the detection of a larger number of compact AGNs in VLBI mode, as well as the successful imaging of Sgr~A* and M87. 

\begin{table*}
\caption{Parameters of Millimetron Space Observatory.}
\label{Table_MM}
\begin{tabular}{lc}
\hline
Parameter					  		 & Value			 \\
\hline
Primary mirror diameter              & 10~m               \\ 
Primary mirror wavefront accuracy    & $<$ 10~$\mu$m RMS  \\
Primary mirror temperature           & $<$ 10~K at L2 orbit and     \\
                                     & $\approx$~70~K at elliptical orbit \\
Orbit                                & Anti-Sun Lagrangian L2 and \\
                                     & High Elliptical near-Earth orbit (HEO) \\
\hline
VLBI subsystem:   	                    &                       	\\
VLBI Band 1 (TBC)                       & 33 - 50~GHz, T$_{sys}<$17~K\\
VLBI Band 2                             & 84 - 116~GHz, T$_{sys}<$37~K\\ 
VLBI Band 3                             & 211 - 275~GHz, T$_{sys}<$50~K\\
VLBI Band 4 (TBC)                       & 602 - 720~GHz, T$_{sys}<$150~K\\
IF bandwidth         		            & 1 -- 2~GHz per 1 channel (up to 4~GHz)  \\
Downlink data rate                      & 1.2~Gbit/s     \\
Time/frequency standard                 & Active Hydrogen Maser \\
On board memory                         & 10 TB -- 100 TB\\
\hline
\end{tabular}
\end{table*}

Millimetron space observatory will operate in halo orbit around L2 point of the Sun-Earth system. Halo orbit is a quasi-stable orbit, located in the vicinity of L2 point in the plane perpendicular to the ecliptic plane. Such orbital configuration is the most suitable for single dish observations in terms of thermal and radiation conditions. L2 orbit will provide the lowest possible temperature of the telescope mirror and thus allow to reach ultimate bolometric sensitivity.

For single dish observations L2 orbit has another advantage -- within half a year, Millimetron will be able to observe practically full celestial sphere. However, for VLBI mode L2 orbits impose strict limitations on the Space-Earth VLBI imaging observations. Feasibility studies of L2 orbits for Space-Earth VLBI showed several significant disadvantages for imaging, as described below in Section~\ref{section:orbit} \citep{Shaykh2020}. To resolve this, it is considered, that Millimetron observatory will be operating also in high elliptical near-Earth orbit after the operation at L2 point. Such configuration allows to achieve better $(u, v)$ coverage for two-dimensional imaging observations. For the presented simulations we selected two high elliptical orbits that have the possibility of transfer from L2 halo orbit and optimized for imaging of Sgr A* and M87.

In contrast to Radioastron, Millimetron will perform Space-Earth VLBI observations without simultaneous data transfer to the ground. Received signal will be digitized and stored in on board memory (10 to 100 Tb, that corresponds to 1.5 -- 15 hours of observations). 
The data transmission to ground can be performed after the observation or in the gaps between the observed scans. Such approach will not limit the observations itself. Expected mission launch date is 2029.

\section{Simulations}\label{siml}
\subsection{Setup}

The main goal of the simulations is to obtain $(u,v)$ coverage for Millimetron Space-Earth VLBI configurations and apply source models of Sgr~A* and M87 to them. Such simulations illustrate  possible advantages that can be reached by S-E VLBI. 

The first part of simulations was devoted to the high resolution imaging of the time averaged models, neglecting the short-time variability of the sources. The second part was devoted to the
observations of Sgr~A* in dynamics over time to study variability of the brightness distribution around the shadow.

The following parameters were used in the simulations: $\Delta\nu=2$~GHz bandwidth, $t=15$ hours of total observing time at central frequency of 230~GHz to be compatible with the EHT. The total duration of observations $t$ was selected according to the parameters of the Millimetron bandwidth $\Delta\nu$ and on board memory capacity (100 Tb). The total volume of on-board memory is limited by mass constraints.

Simulations consisted of several steps: 
\begin{itemize}
    \item Calculation of $(u, v)$ coverages, taking into account the source visibility for space and ground telescopes. 
    \item Calculation of interferometric visibility functions for the corresponding $(u, v)$ coverages using the specified source models. 
    \item Application of sensitivity, phase errors and noise.
    \item Obtaining initial dirty map. 
    \item Image reconstruction.
\end{itemize}

Calculations of the $(u, v)$ coverage and the model application, as well as the
image reconstruction were performed using Astro Space Locator Software (ASL). This software package is one of the results of Radioastron mission development that included the establishment of data processing pipeline and software for post-correlation data analysis \citep{ASCCorrelator, ASLCASA}. During the mission operations this software was verified and used for fringe search, fringe fitting, imaging and simulations of VLBI observations.
For ground support of S-E VLBI simulations we took the telescopes of EHT collaboration. The coordinates and parameters of selected telescopes used in our simulations are provided in Table \ref{Table_EHT}).

Before proceeding with VLBI imaging simulations we have compared ASL performance with EHT data processing pipeline that uses Maximum Entropy Method (MEM) imaging algorithms (\cite{EHT2019_p6}, Section 6.2.2). And in order to track down the impact of the space-ground baselines, we performed the simulations for pure ground baselines and space-ground baselines.

Orbital calculations were performed with the software developed for Millimetron mission scheduling and flight dynamics calculations at Astro Space Center of Lebedev Physical Institute. 

We took the duration of single observing segment according to the coherent integration time of 10~s. This value is a conservative estimate for the atmospheric coherence time at ground-based EHT sites. These limitations were taken into account at  sensitivity application step.

For Sgr~A* we performed the simulations with the $(u, v)$ coverage accumulated for 10 days, which corresponds to one orbit period, i. e. 15 hours or 5400 observing segments were distributed in equally across the 10 days. For dynamic imaging we calculated set of $(u, v)$ coverages each corresponding to a single frame in the perigee of orbit.

For M87 we selected a 20 hour region of the orbit, that covers the baselines from 0.5 up to 6 Earth diameters. In this case the total number of single observing segments was distributed across 20 hours correspondingly.

Due to the fact that Sgr~A* and M87 are located in different parts of the celestial sphere, a highly elliptical orbit with a given energy can be oriented in such a way as to provide the best (u,v) coverage over the entire period (about 9-10 days) for only one source. At the same time, such orbit is capable to observe the second source with baseline projection less than 5 Earth diameters, only in the perigee, that is about 20 hours.

The sensitivity for each baseline was calculated using available ground telescope system equivalent flux density ($SEFD$) (see Table \ref{Table_EHT}) \citep{Chael2016, Chael2018, 2019ApJ...875L...2E}. For Millimetron the estimated sensitivity ($SEFD$) at 230~GHz is $\sim 4000$~Jy.

Simulations of baseline and telescope sensitivity in Astro Space Locator were implemented as follows: a vector with an amplitude of sensitivity (calculated from the values of $SEFD$, bandwidth and integration time) and a random phase noise were added to each value of the visibility function in the resulting synthetically simulated observations.

For simplicity, the loss of absolute phase has not been considered in this paper. The simulations were done using synthetic data that assumes accurately calibrated phases on the ground-ground baselines. We believe that this error does not fundamentally affect image quality, because in this simulation we discuss Millimetron capabilities in HEO orbit. Such orbital configuration generate enough closure phase triangles for imaging. While we neglect phase errors in this paper, in practice imaging should be done either with iterative CLEAN + self-calibration or directly  with closure phases.

In order to perform a quantitative evaluation of image quality we used fidelity measure:
\begin{equation}\label{eq:fidelity} 
F = \frac{MAX(M_i)}{\sqrt{\frac{1}{n}\sum_{i=1}^{n}(I_i - M_i)^2}}
\end{equation}
Where $F$ is fidelity, $M_i$ is the intensity at the $i$ pixel in the model image, $I_i$ is the intensity at the $i$ pixel in the reconstructed image, $n$ is a number of pixels in the image.
\begin{table*}
\caption{\textbf{Parameters of ground telescopes at 230 GHz \citep{2019ApJ...875L...2E}}}
\resizebox{\linewidth}{!}{%
\begin{tabular}{lrrrrr}
\hline
Telescope												&X,~m		&Y,~m			&Z,~m		&SEFD,~Jy		&$D$,~m	\\
\hline
Atacama Large Millimeter Array, Atacama, Chile (ALMA)       					&  2225061.164  	&   -5440057.37    		&   -2481681.15    	&     74 		& 73      	\\
Atacama Pathfinder Experiment, Atacama, Chile (APEX)					&  2225039.53   	&   -5441197.63    		&   -2479303.36    	&   4700 		& 12      	\\
Greenland Telescope, Greenland (GLT)   						&  1500692.00   	&   -1191735.0     		&    6066409.0     	&  5000  		& 12      	\\
IRAM 30-m millimeter radio telescope, Pico Veleta, Spain (PV)					&  5088967.900	&    -301681.6000  		&    3825015.8000  	&   1900 		& 30      	\\
James Clerk Maxwell Telescope, Hawaii (JCMT)						& -5464584.68   	&   -2493001.17    		&    2150653.98    	&  10500 		& 15      	\\
Large Millimeter Telescope, Mexico (LMT)							&  -768713.9637 	&   -5988541.7982  	&    2063275.9472 	&   4500 		& 50 		\\
Submillimeter Telescope, Arizona, United States (SMT)					& -1828796.200  	&   -5054406.800   		&    3427865.200   	&  17100		& 10      	\\
Submillimeter Array, Hawaii, (SMA)								& -5464523.400  	&   -2493147.080   		&    2150611.750   	&   6200 		& 14.7      	\\
\hline
Kitt Peak National Observatory, Arizona, United States, (KP)$^{2020}$			& -1995678.840  	&   -5037317.697   		&    3357328.025   &  13000 		& 12     	\\
Northern Extended Millimeter Array, Plateau de Bure, France (NOEMA)$^{2020}$ 	&  4523998.40	&     468045.240  		&    4460309.760   &  700  		& 52      	\\
\hline
\end{tabular} 
}
\label{Table_EHT}
\raggedright
$^{2020}$ \small{- telescopes to be added to the EHT in 2020}.
\end{table*}

\subsection{Orbit Configurations}
\label{section:orbit}
Halo orbit in L2 point of the Sun-Earth system provides the best environment for the single dish observations. At the same time, it lacks short and intermediate baseline projections for two-dimensional VLBI imaging. Shorter projections for space-ground baselines provide less gaps in the $(u, v)$ coverage and give intersections between the solely ground and space-ground baselines. Having smaller gaps in the $(u, v)$ coverage is crucial for the quality of the reconstructed image. Moreover, the period of the halo orbit is half a year, so that for given sources short baseline projections will be available only once per year for short time spans (up to 2~days). This makes difficult to obtain an acceptable $(u, v)$ coverage and perform regular imaging VLBI observations with Millimetron in L2 point. 

For that reason we considered the possibility of Millimetron to operate at near-Earth orbit. The choice fell on high elliptical orbits (HEO) with apogee close to the Moon's orbit, because these types of orbits can provide transfer from L2 with gravity assist around the Moon \citep{GA_HEO}. Such maneuver can save the $\Delta v$ (momentum) budget for the further mission stages. 

For orbital calculations we used Prince-Dormand propagator of the 4th-5th orders with adaptive step size \citep{PD45} as it provides a good performance for HEO. The force model included 4 harmonics of the EGM96 Earth gravity model \citep{EGM96} and perturbations from the Moon point of mass. The ephemeris of the Moon was calculated using DE431 \citep{Folkner2014}.

The following orbital constraints were considered: distance at perigee $r_{p}\geq 10000$ km and the distance at apogee $r_{a}\geq340000$ km. The goal was to obtain relatively compact $(u, v)$ coverage within several Earth diameters (ED). 

The distribution of the points in the $(u, v)$ plane depends on the semi-major axis, because it determines the lower limit of baseline projections. Upper boundaries for baseline projections were set as $\leq5$ ED, which corresponds to the angular resolution of $\sim 4$~$\mu$as. Thus, the semi-minor axis was set to $b=\sqrt{r_{a}r_{p}}\geq56000$ km.

\begin{figure*}
\begin{center}
\includegraphics[width=\textwidth,angle=0]{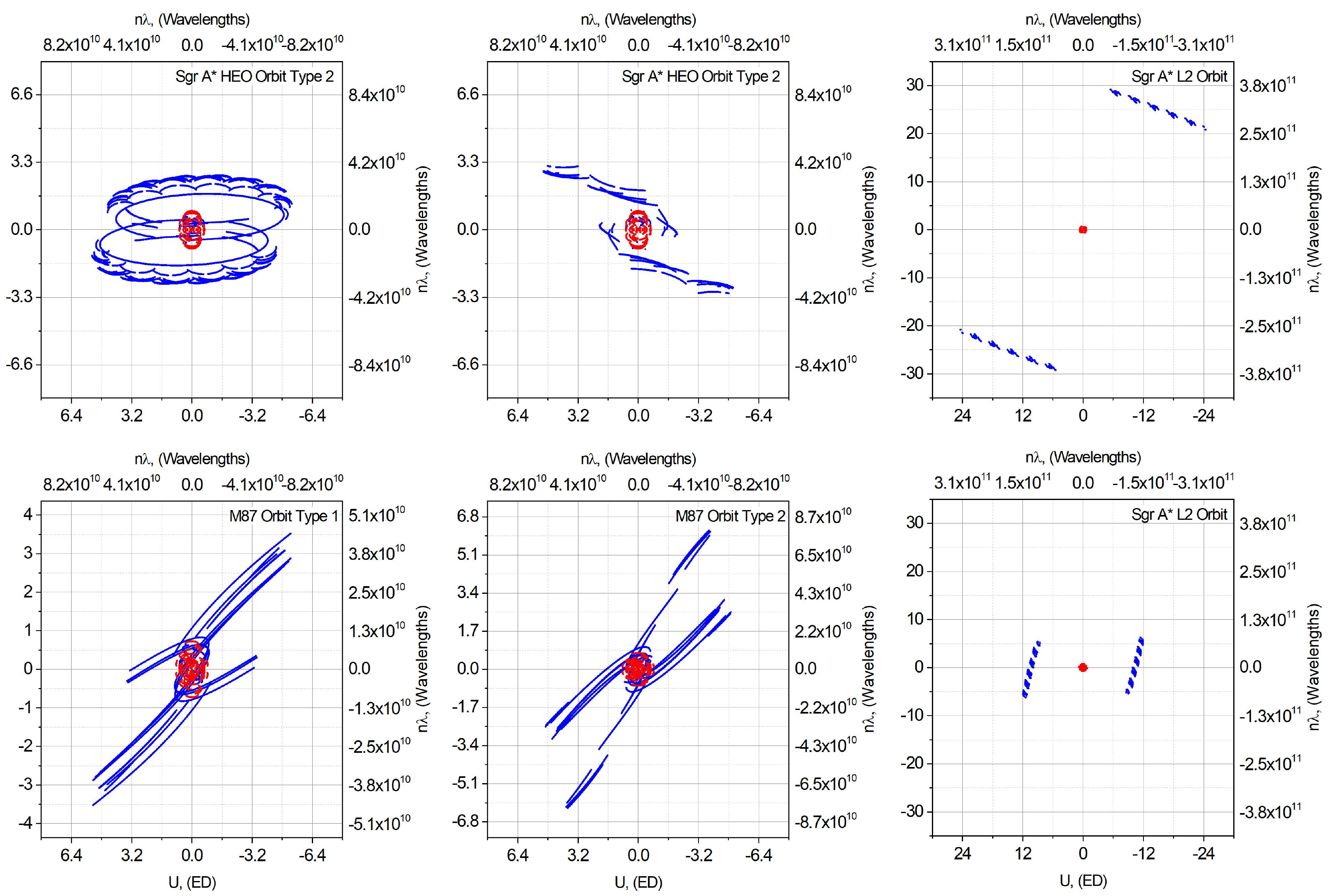}
\end{center}
\caption{Top: $(u, v)$ coverage for Sgr~A* (from left to right: Orbit Type 1, Orbit Type 2 and L2). Bottom: $(u, v)$ coverages for M87 (from left to right: Orbit Type 2 and L2). Coordinates are represented in Earth diameters (bottom-X, left-Y axes) and wavelengths (top-X, right-Y axes). Red dots correspond to EHT ground baselines, blue dots correspond to EHT+MM space-ground baselines}.
\label{fig:simuv}
\end{figure*}

The calculations resulted into the set of orbits. Among this set, two orbits were selected to have the minimal amount of gaps in the $(u, v)$ coverage for Sgr A* (Orbit type 1, see Fig.~\ref{fig:simuv}, top) and for Sgr A* and M87 (Orbit type 2, see Fig.~\ref{fig:simuv}, bottom). Parameters of selected orbits are shown in the Table~\ref{Table_orbits}.
\begin{table}
\small
\begin{center}
\caption{Orbit parameters}
\begin{tabular}{lrr}
\hline
Parameter & Type 1 & Type 2 \\
\hline
$a$      & 165,000~km       & 165,000~km  \\
$e$      & 0.939           & 0.939 \\
$i$      &20.008           & 323\\
$\Omega$ &$-3.583^{\circ}$ & $170^{\circ}$\\
$\omega$ &$-92^{\circ}$    & $-114^{\circ}$\\
Period of both orbits is 10\\
\hline
\end{tabular}
\label{Table_orbits}
\end{center}
$a$ -- semi-major axis, $e$ -- orbit eccentricity, $i$ -- inclination, $\Omega$ -- longitude of the ascending node, $\omega$ -- argument of periapsis, i. e. the orientation of the ellipse in the orbital plane
\end{table}

\begin{figure*}
\begin{center}
\includegraphics[width=\textwidth,angle=0]{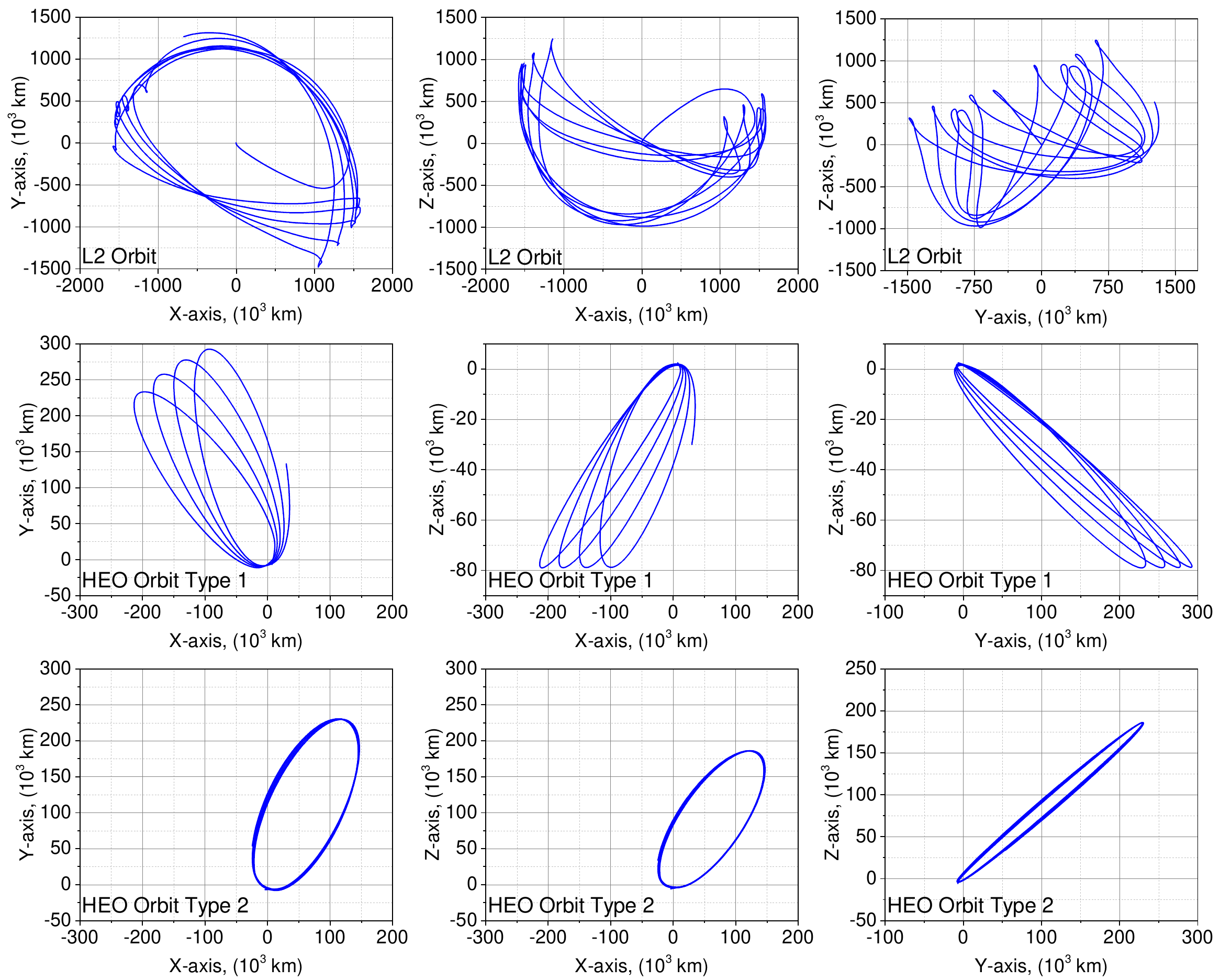}
\end{center}
\caption{Projection of considered orbital configurations in geocentric coordinate system. Top panel: L2 Orbit, middle panel: HEO Orbit Type 1, bottom panel: HEO Orbit Type 2. Coordinates are represented in thousand kilometers.}
\label{fig:p_orbits}
\end{figure*}

\begin{figure*}
\begin{center}
\includegraphics*[width=\textwidth,angle=0]{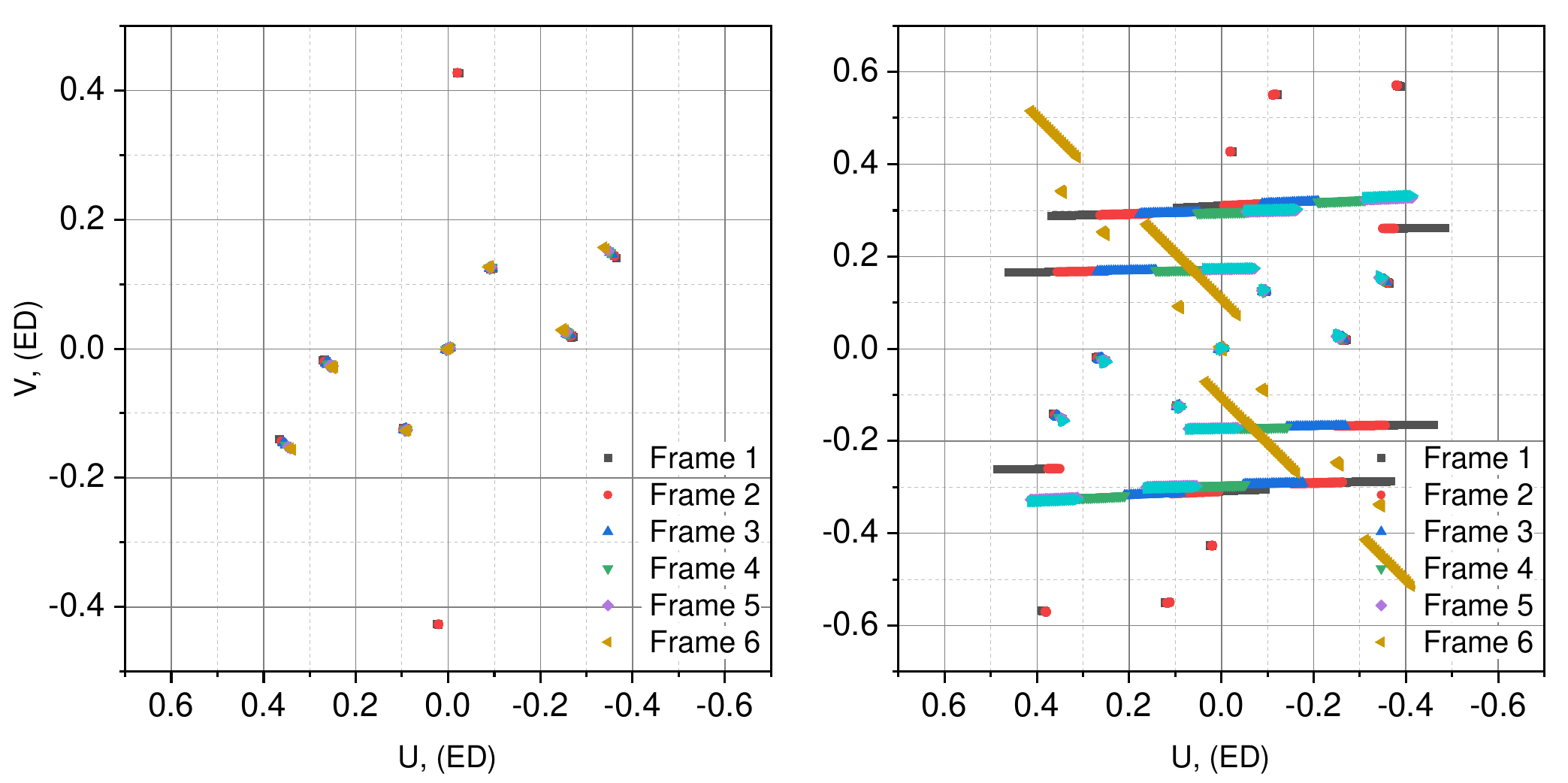}
\end{center}
\caption{$(u, v)$ coverage for Sgr A* dynamic simulations: EHT only (left) and Millimetron+EHT (right, Orbit Type 1) dynamic frames. Coordinates are represented in Earth diameters. The total snapshot integration time is 1326 s (221 s per frame).}
\label{fig:movieuv}
\end{figure*}

Orbit type 2 in Table~\ref{Table_orbits} was calculated to provide possible optimal imaging observations for two sources with acceptable $(u, v)$ coverage. It was assumed, that observations of M87 and Sgr~A* will take place at different orbit sections. The latter orbit was used in the simulations for M87.

It is supposed, that Millimetron will operate at near-Earth orbit at least several years. The range of tasks to be solved will not be limited by imaging of the two indicated sources and it is planned to conduct observations of other compact extragalactic sources.

\subsection{Source Models}\label{source}
In our simulations we used averaged models described in \citep{Moscibrodzka2014} for Sgr~A* and kindly provided for us by M. Mos\'cibrodzka. They include a set of time-averaged (over $\Delta t\approx 3$ hours) models: \# 16, \# 24, \# 31 and \# 39 in nomenclature of \citep[][Table 1]{Moscibrodzka2014}, and are shown in Fig. \ref{fig:sag_ehtmm} (left column). Models differ by the inclination angle $i$ between the BH spin and the observer's line of sight, the electron temperature $\Theta_{e,\rm j}=k_BT_e/m_ec^2$ in jet, the ratio of protons to electrons temperatures $T_p/T_e$, and the accretion rate $\dot M$ are shown in Table \ref{Table_mod}. The models have chosen such to emphasize their characteristic peculiarities, in order to better reveal the differences between imaging with EHT-only and EHT+Millimetron. Both diffractive and refractive scattering were included in the simulations for Sgr~A* using the parameters constrained in \citep{Scattering}.

It seems obvious and is confirmed both in observations and in numerical simulations, that the flow of innermost accretion and its emissivity are time dependent, and such are also their images around the shadow as well. Near-infrared (NIR) monitoring of Sgr~A* reveals variability on a wide range of time scales from 20~s to hours with variations of magnitude up to factor of 10 \citep{Witzel2018,Chen2019}. Besides such regular oscillations of disc emissivity, strong sporadic flares can also happen in the innermost region, such as, e.g., the ``non-thermal bomb'' in May 2019 detected in NIR at the Keck Telescope \citep{Do2019} and occurred apparently within $\sim 10r_g$ of the innermost accretion flow \citep{Gutierrez2020}. It is quite expected that such regular variability and explosive events might be manifested in sub-mm range on similar time scales. Therefore, dynamic interferometry is essential to understand the physical processes in this area, and moreover to monitor effects of variability on to the image itself. EHT+Millimetron capabilities are sufficient to perform observations of such variability with proper time resolution.

For dynamic simulations of Sgr~A* we used the model \# 39 (Fig. \ref{sag_dyn}, left column) from Table \ref{Table_mod}. According to the modeling parameters, the time spacing between the GRMHD movie frames is $221$~s. It is 10 times of the gravitational timescale $t_{g}=GM/c^{3}$.

\begin{table}
\begin{center}
\caption{List of EHT models used for S-E VLBI simulations.}
\label{Table_mod}
\begin{tabular}{l|l|l|l|l}
\hline 
\#			    & $i$         & $\Theta_{e,\rm j}$   &  $(T_p/T_e)_{\rm d}$       &  $\dot M~(M_\odot~{\rm yr}^{-1})$			 \\
\hline
16              & 60$^\circ$         &    10       &       5       &     $3.9\times 10^{-9}$             \\ 
24              & 60$^\circ$         &    20       &       20        &     $4.2\times 10^{-8}$           \\ 
31              & 30$^\circ$         &    10       &       5       &     $5.6\times 10^{-9}$             \\ 
39              & 30$^\circ$         &    20       &       20       &     $4.1\times 10^{-8}$            \\
\hline
\end{tabular}
\\
\end{center}
$i$ is the inclination of the BH spin to the observer's line of sight, $\Theta_{e,\rm j}$ is the electron temperature (in $m_ec^2$) in the jet, 
$(T_p/T_e)_{\rm d}$, the ratio of proton to electron temperature in disk, $\dot M$, the accretion rate, the nomenclature of models is as in Table \ref{Table_EHT} in 
\citet{Moscibrodzka2014}. 
\end{table}

In our simulations of the ring around the shadow of M87 we have used the model described by \cite{johnson2019universal}. It is a time-averaged image of GRMHD simulation of M87 with parameters chosen to be consistent with the EHT data of 2017 (\cite{2019ApJ...875L...5E}) as specified in \citep{johnson2019universal}. The parameters are as follows: $M = 6.2 \times 10^9 M_\odot$, $a/M = 0.94$, $i = 163^\circ$ and rate of mass accretion is matching the flux density of EHT 2017 observations at 1.3 mm. Fig. \ref{fig:m87} (a) shows the model.

\section{Results}\label{res}

\subsection{Static Images}
\label{sec:simres}
\begin{figure*}
\begin{center}
\includegraphics[width=\textwidth,angle=0]{{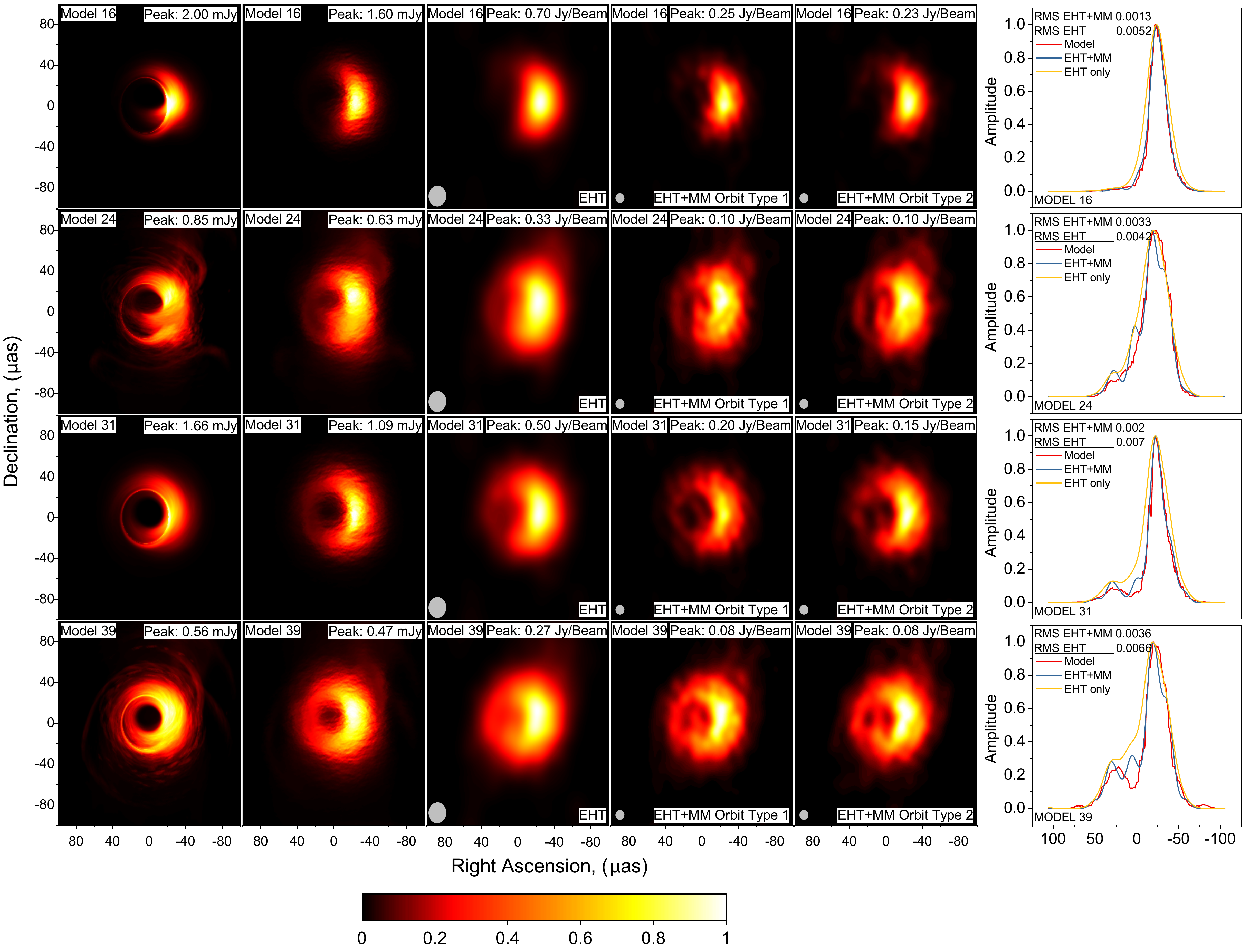}}
\end{center}
\caption{Results of simulations for Sgr~A* static imaging at 230 GHz: initial models unscattered (leftmost) and scattered (left), CLEAN images for EHT only (left middle), CLEAN images for EHT+Millimetron Orbit Type 1 (right middle), CLEAN images for EHT+Millimetron Orbit Type 2 (right). From top to bottom: model \# 16, model \# 24, model \# 31, model \# 39. Resulting image size: 200 $\times$ 200 $\mu$as. The rightmost panel depicts the brightness profiles as drawn from the origin $(x,y)=(0.0)$ along the positive direction of $x$-axis; RMS for EHT-only and MM+EHT are shown in the legends left-uppermost corner of each plot. The size of synthesized beam is 22~$\times$~19~$\mu$as at -10.53$^{\circ}$ for EHT only configuration and 10~$\times$~10~$\mu$as at 9.84$^{\circ}$ for Millimetron+EHT. Color scale is linear multiplied by peak value of corresponding panels.}
\label{fig:sag_ehtmm}
\end{figure*}

Fig.~\ref{fig:sag_ehtmm} show CLEAN images of Sgr~A* for EHT only/Millimetron+EHT VLBI configurations. 

\begin{figure*}
\begin{center}
\includegraphics[width=\textwidth,angle=0]{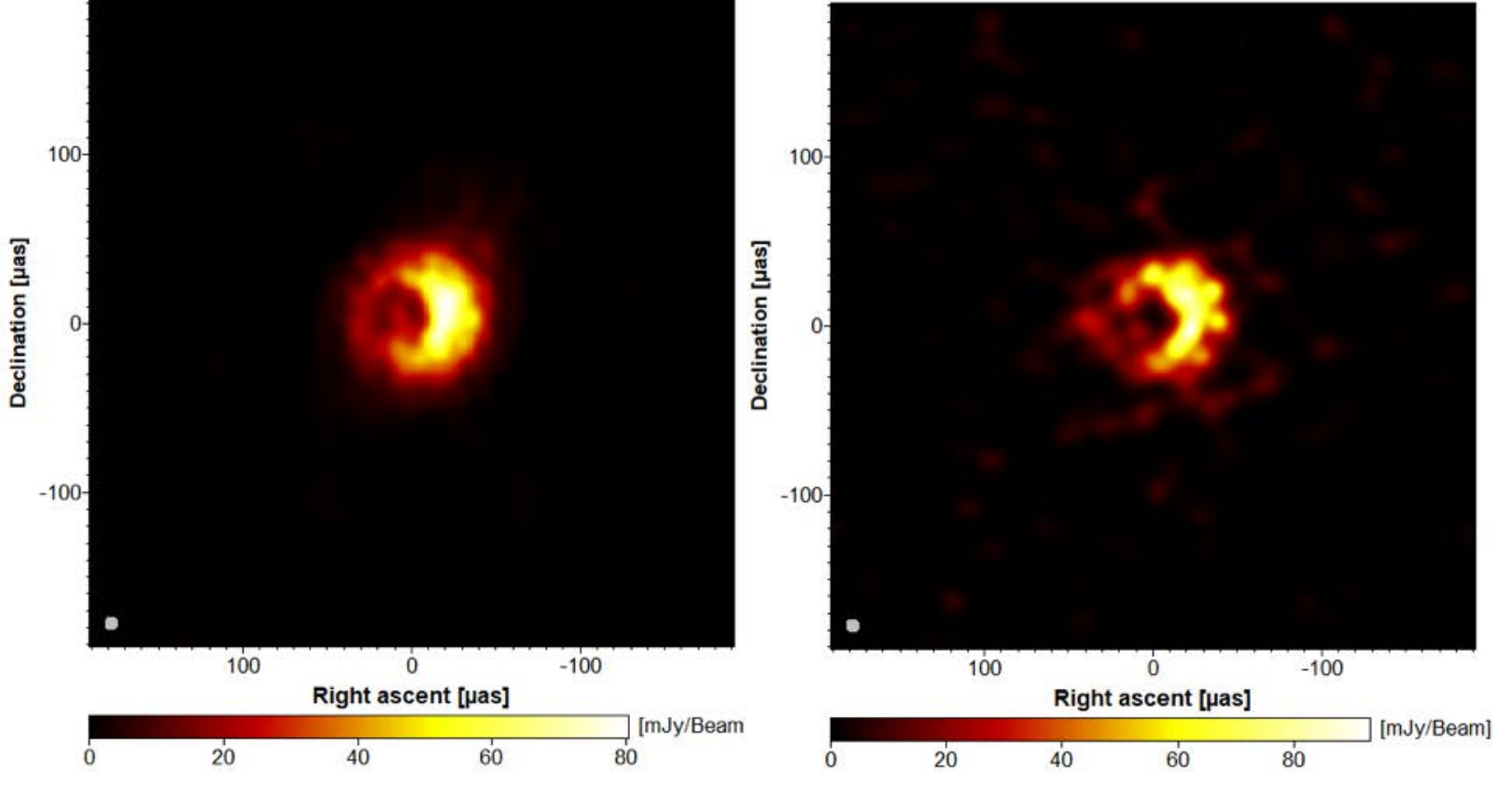}
\end{center}
\caption{Left plot corresponds to imaging with Sgr A* \# 39 model (averaged), right plot corresponds to imaging with Sgr A* \# 39 dynamic model (cycled 80 different frames with time separation 221 s, and the concerted (u,v)-plane).}
\label{fig:fig_uv_cycled}
\end{figure*}

The models of Sgr A* that were used in simulations (see Section \ref{source}) are averaged. These time-averaged images represent an idealized scenario for Sgr A* due to GRHMD simulations predict high time variability in the source.
Of course, imaging the source will be much harder for Sgr A* when the underlying data is evolving in time. In the paper, we concentrate on the $(u,v)$-coverage quality anticipated for the HEO orbit. We assumed that after 10 days of integration, we will get a certain average image that will be in a good correspondence with averaged GRMHD model. Fig.~\ref{fig:fig_uv_cycled} demonstrate correspondence between static image reconstruction of averaged model (left panel) and dynamic image reconstruction when the underlying data is evolving in time (right panel). 
From this figure it can be seen that the resulting images are quite similar, so further we will use this assumption.  An additional reason of using the averaged model instead of the dynamic one is that the averaged model is better for fidelity computation, and simpler for comparison between model images and those obtainable with synthetic observations.

The quality of images obtained by EHT and EHT+MM observations can be quantified by a characteristic similar to the sharpness of the inner edge defined in \citep[][see Eq. 22 and 23]{EHT2019_p6} as the ratio of the smoothing kernel width to the crescent diameter $s=w_s/d_c$. The plots of normalized brightness profiles shown on the right panel in Fig.~\ref{fig:sag_ehtmm} illustrate differences in shadow sharpness obtained with EHT-only and with Millimetron+EHT VLBI. The differences as seen from comparison of the profiles along the $x$-axis for the model, EHT and Millimetron+EHT are clearly smaller in the latter case. Their numerical values are indicated in Table \ref{Table_RMS_Sharp}. EHT+MM synthetic observations show a factor of $\sim~2$ better correspondence to the model sharpness. Additionally to avoid the effects of Doppler beaming at high inclination angles on to RMS we have analyzed separately the left part of models and obtained images profiles (see Table \ref{Table_RMS_Sharp}, two last columns). EHT+MM shows better correspondence here as well, giving a more detailed imaging of fainter parts of the initial model.

Another characteristic is the flux depression, i.e, the ratio of the floor to the average over crescent brightness $f_d=F_f/\langle F_c\rangle$ can be also roughly estimated from observations of brightness profiles in Fig.~\ref{fig:sag_ehtmm}: in all cases brightness in the local minimum at the origin is shown by MM+EHT images are lower than in EHT-only ones, in models 31 and 39 this difference in depressions is around factor 2. 

\begin{table*}
\small
\begin{center}
\caption{Sharpness: RMS for EHT only and Millimetron+EHT configurations}
\begin{tabular}{lcccr}
\hline
\#    & EHT-Model   & (EHT+MM)-Model & EHT-Model & (EHT+MM left side)-Model \\
\hline
16      & 0.0013       & 0.0052  &  0.0022      & 0.00069\\
24      & 0.0033       & 0.0042  &  0.0049      & 0.0045\\
31      & 0.002        & 0.007   &  0.0049      & 0.0019\\
39      & 0.0036       & 0.0068  &  0.0088      & 0.0052\\
\hline
\end{tabular}
\label{Table_RMS_Sharp}
\end{center}
\end{table*}

\begin{figure*}
\begin{center}
\includegraphics[width=\textwidth,angle=0]{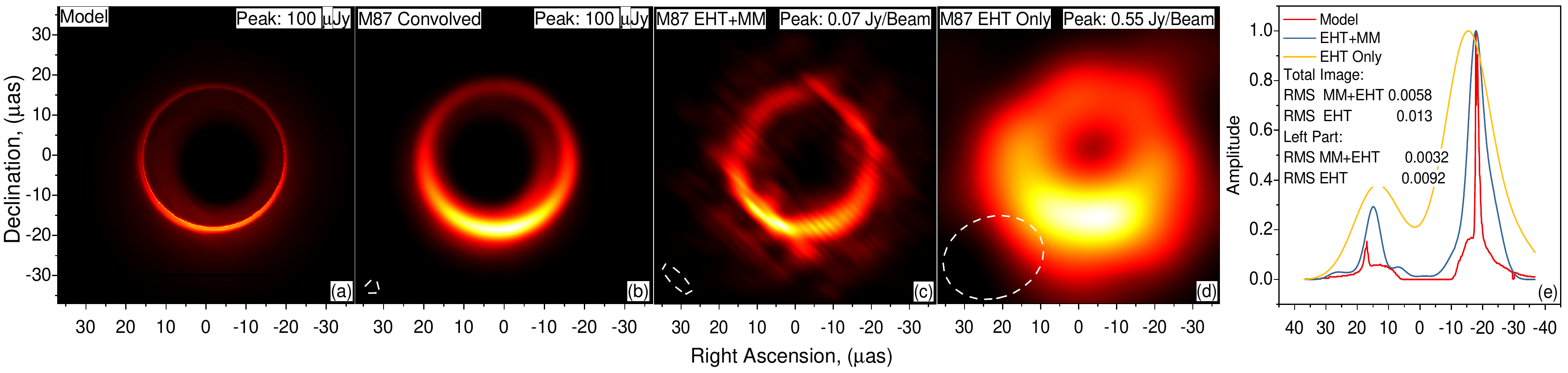}
\end{center}
\caption{Results of simulations for M87 (Orbit Type 2) at 230 GHz: initial convolved models (left), CLEAN images for EHT+Millimetron (middle), CLEAN images for EHT only (right). Images size: 72~$\times$~72~$\mu$as. White dashed lines show the synthesized beam, the size of beam is 17.5~$\times$~15.8~$\mu$as at -71.56$^{\circ}$ for EHT only configuration and 5 $\times$ 5 $\mu$as at 46.98$^{\circ}$ for EHT+Millimetron.  Color scale is linear multiplied by peak value of corresponding panels.}
\label{fig:m87}
\end{figure*}

In order to characterize quality of model image recovery we use normalized image fidelity and SSIM index \citep{Wang2004}. Normalized fidelity is given by (\ref{eq:fidelity}) where both $I_i$ and $M_i$ images are normalized to 1. This represents synthesized image characteristics with respect to the original model in terms of mean square difference. If images match perfectly then the fidelity is infinite. The SSIM index correlates well with human perception of the image quality and used here as human independent measure of this parameter \citep{Wang2004}. The SSIM index value ranges from 0 to 1. SSIM index value 1 is for two similar images. 
As reference for comparison we used the original model image of M87 as in Fig.~\ref{fig:m87} (a) and model image which is convolved with a circular Gaussian distribution of an angular size corresponding to the largest angular dimension of the synthesized beam for EHT+Millimetron image simulation see Fig.~\ref{fig:m87} (b). The convolved image represents measurements of a model with an ideal diffraction limited telescope of an aperture size of order of Millimetron orbit. Comparison with this image allows to analyze relative amount of image artefacts introduced due to particular $(u, v)$ coverage.
Table~\ref{tab:fidelity} shows fidelity and SSIM values for Millimetron+EHT and EHT only images using M87 model and convolved M87 model as shown in Fig.~\ref{fig:m87} (a-d). Both image quality measure demonstrate significant, up to 10 times, improvement of image reconstruction for Millimetron+EHT in comparison with EHT only. Especially clear difference is seen when convolved image is used for comparison. It must be noted that Millimetron+EHT has a higher normalized fidelity than convolved model which reflects the fact that convolution beam is a circular beam with an average size derived from the synthesized beam of EHT+Millimetron $(u,v)$ coverage which is elliptical. As a result Fig.~\ref{fig:m87} (c) contains more spatial information than Fig.~\ref{fig:m87} (b).

\begin{table}
\begin{center}
    \caption{Normalized Fidelity and SSIM for images of M87}
    \label{tab:ssim}
    \begin{tabular}{p{1.4 cm}|p{1.5cm}|p{2.6cm}|p{1cm}}
        \hline
        Model  & Convolved & Millimetron + EHT &  EHT \\
        \hline
        & \multicolumn{3}{|c|}{Normalized Fidelity} \\
        \hline
        Model   & 9.79 & 12.56 & 2.25 \\
        Convolved & $\infty$ & 22.40 & 2.55 \\
        \hline
         & \multicolumn{3}{|c|}{SSIM} \\
        \hline
        Model   & 0.286 & 0.345 & 0.012    \\
        Convolved & 1 & 0.93 & 0.126  \\
        \hline
    \end{tabular}
\end{center}
\end{table}

Fig.~\ref{fig:m87} shows VLBI images of the photon ring model for M87 (model is on the left of Fig.~ \ref{fig:m87}) described by \citet{johnson2019universal}, for EHT-only (middle) and EHT+MM (right) synthetic observations.

\begin{table}
\begin{center}
    \caption{Fidelity of obtained images for Sgr A*.}
    \label{tab:fidelity}
    \begin{tabular}{c|c|c|c|c}
        \hline
        \multicolumn{5}{|c|}{Orbit Type 1} \\
        Sgr A* Model    & 16 & 24 & 31 & 39 \\
        \hline
        MM+EHT          & 61.39 & 49.49 & 54.18 & 41.2\\
        EHT only        & 32.53 & 34.65 & 31.52 & 29.1\\
        \hline
        \multicolumn{5}{|c|}{Orbit Type 2} \\
        \hline
        MM+EHT      & 59.58 & 45.77 & 52.14 & 37.97\\
        EHT only    & 31.44 & 33.35 & 30.57 & 28.28\\
        \hline
    \end{tabular}
\end{center}
\end{table}

The asymmetry of the images is related to the magnitude of the black hole spin and the angle between the axis of rotation of the black hole and the direction towards the observer. Position of the center of mass for the M87 source determined from Fig. \ref{fig:m87} are presented in Table \ref{tab:cm_m87}. The accuracy of asymmetry parameters determination is higher for joint EHT and Millimetron observations.

To estimate the degree of asymmetry of the images, we calculated the position of the center of mass of each image:

\begin{eqnarray}\label{eq:m_c} 
X_{cm} = \frac{\sum_{i = 1}^{N} I_i \cdot x_i} { \sum_{i = 1}^{N} I_i }  \\
Y_{cm} = \frac{\sum_{i = 1}^{N} I_i \cdot y_i} { \sum_{i = 1}^{N} I_i }  \nonumber \\
R_{cm} = \sqrt{X_{cm}^2 +Y_{cm}^2} \nonumber \\
\varphi_{cm} = \arctan{(Y_{cm} / X_{cm})} \nonumber 
\end{eqnarray}

Here $x_i$, $y_i$ are the coordinates of each image pixel, $I_i$ is the intensity in image pixel and $N$ is a number of image pixels.

\begin{table}
\begin{center}
    \caption{Center of mass position for M87.}
    \label{tab:cm_m87}
    \begin{tabular}{c|c|c|c}
              & Model & EHT only & EHT + Millimetron  \\
        \hline
        $\varphi$    & -107.2 & -112.0 & -111.2  \\
        $R$    & 7.8 & 6.9 & 8.0  \\
        \hline
    \end{tabular}
\end{center}
\end{table}

It is more noticeable on the M87 example that introducing the space element to the VLBI configuration provides a higher resolution leading to the finer structure observed in the images. Lack of scattering in the M87 galaxy clearly reveals the advantages from the Space-Earth VLBI. 

\subsection{Dynamic Images of Sgr A*}
The second part of the simulations was devoted to the dynamic imaging. Millimetron spacecraft fly by time in perigee of HEO is about 20 minutes. At this portion of orbit the baseline projection changes within the interval of 0.1 -- 1 Earth diameters resulting to fast $(u, v)$ coverage evolution providing a possibility to carry out a dynamical imaging of Sgr A*.

According to the source model parameters, the time interval between the modelled movie frames for Sgr~A* is $221$ s, i.e. $\approx10\times t_{g}$. Thus, we made and attempt to simulate a short observations of the dynamic imaging in the HEO perigee.

Fig. \ref{fig:movieuv} show the $(u, v)$ coverage for Millimetron+EHT (left) and EHT only (right) dynamic simulation frames. Obviously, the $(u, v)$ coverage of the ground only VLBI configuration remains practically constant within such short time intervals and dynamic imaging is not possible.

\begin{figure*}
\begin{center}
\includegraphics[width=0.55\textwidth,angle=0]{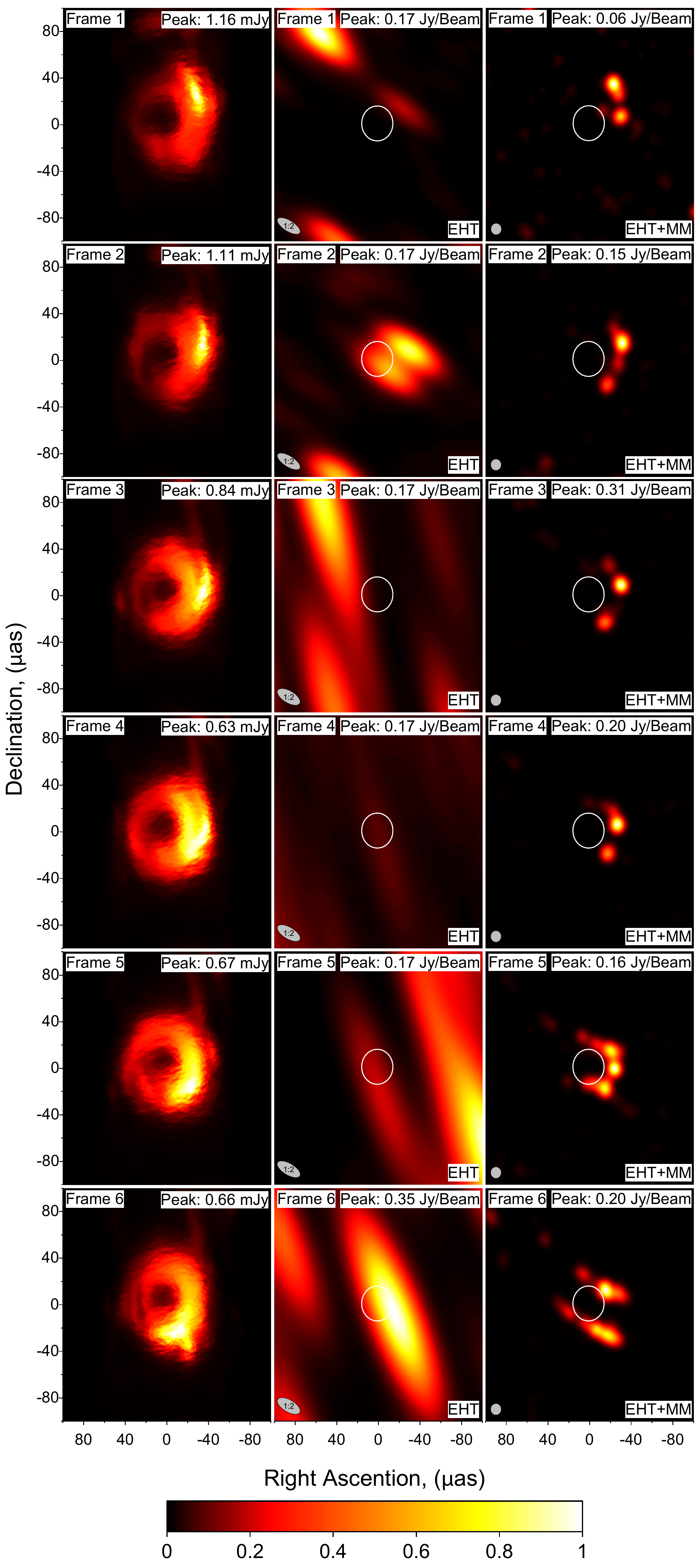}
\end{center}
\caption{Dynamic simulations for Sgr~A* (Orbit Type 1): snapshots of dynamic models (left), CLEAN images for EHT only configuration (middle), CLEAN images for EHT+Millimetron (right). Frames advance from top to with time difference between the frames 221~s. Resulting images size: 200 $\times$ 200 $\mu$as.The white circle in the center of each image indicates (0, 0) coordinates of the source. Color scale is linear multiplied by peak value of corresponding panels.}
\label{sag_dyn}
\end{figure*}

Resulting dynamic images for these simulations are represented on the Fig.~\ref{sag_dyn}. The white circle in the center of each image indicates (0,0) coordinates of the source.
The time difference between the obtained images is 221~s. In other words, each single image corresponds to given 221~s frame of the dynamic model.

The variations of brightness asymmetry around the shadow with characteristic time $t\sim 10t_g$ are clearly seen in Fig. \ref{sag_dyn} (left column), as well as a varying flow in the contiguous parts of the disc. When averaged these details merge into a heavily smeared image as seen in Fig. \ref{fig:sag_ehtmm} (left column).

\begin{figure*}
\begin{center}
\includegraphics*[width=0.47\textwidth,angle=0]{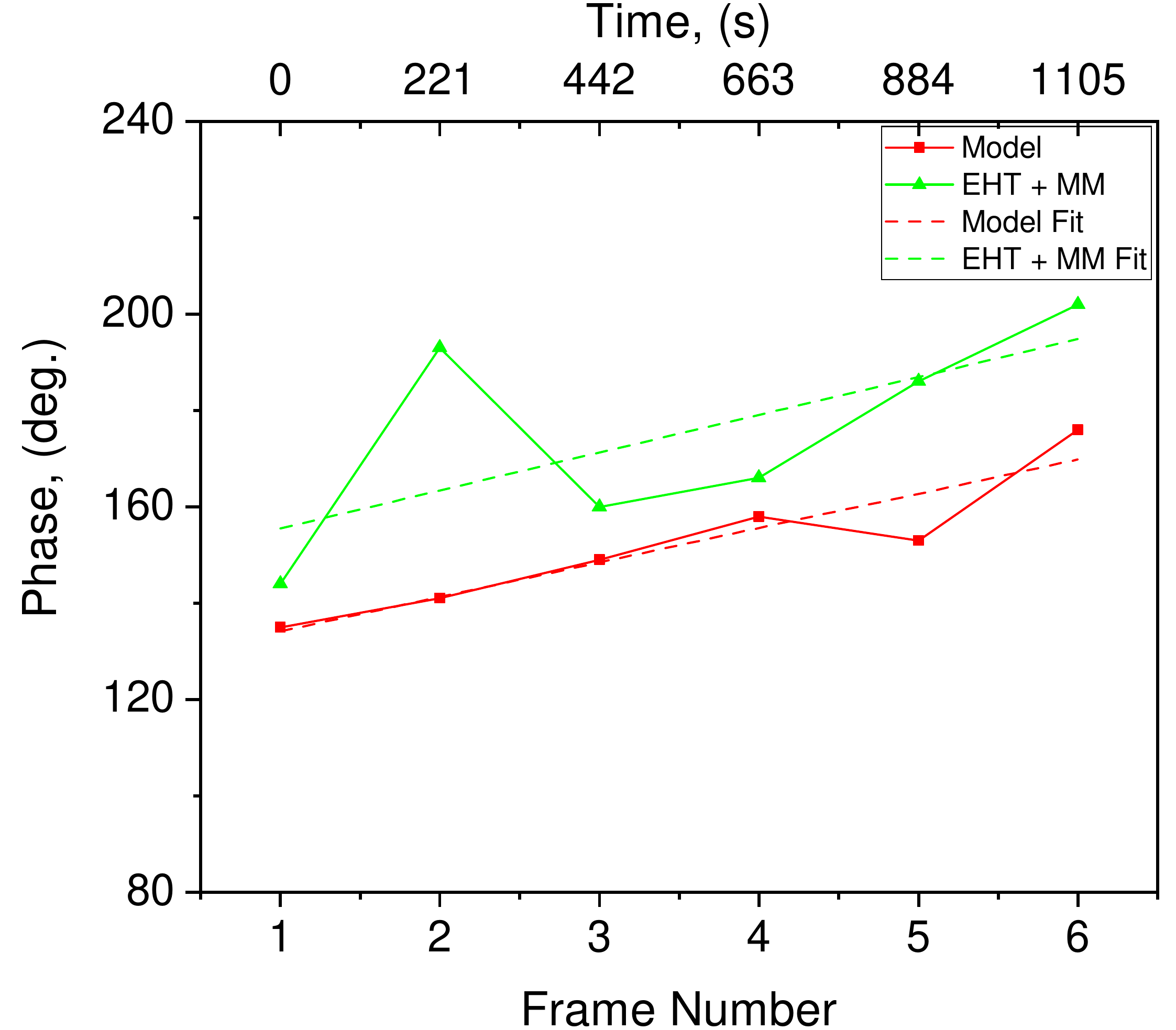}
\end{center}
\caption{Estimation of asymmetry for dynamic imaging: left panel shows the position of center mass along the phase $\varphi$ of Sgr A* dynamic images. Solid lines corresponds to measured values, dash lines corresponds to linear fit. Right panel shows the position of center mass along the radius $R$ of Sgr A* dynamic images.}
\label{fig:cm_sgr_frames}
\end{figure*}

We have estimated the position of center of mass for each frame of dynamic images of Sgr~A*. The results are presented in Fig. \ref{fig:cm_sgr_frames}. For EHT+Millimetron case, the position of center of mass coincides with the models, while for EHT-only case we have essentially random values.

Synthetic images are seen to turn their angular brightness distribution clockwise with time evolution, what may correspond to strong variations of brightness in the accretion disc comparable to the relativistic beaming around the shadow. This turn results in variations of visible asymmetry which can be distinguished by measurements of the instant brightness-weighed angle in the crescent. The estimated rotation velocity of this angle is given in Table \ref{tab:cm_angular_velocity}.

From the slope of the fitted line in Fig. \ref{fig:cm_sgr_frames} (left), we estimated the angular velocity of bright spot rotation in the Sgr~A* model. The results are presented in Table \ref{tab:cm_angular_velocity}. These results show that the quality of the reconstructed dynamic images EHT+MM is sufficient to determine the physical parameters from the image.

\begin{table}
\begin{center}
    \caption{Angular velocity of a bright spot rotation in the Sgr A* model ($ 10^{-4} s^{-1}$).}
    \label{tab:cm_angular_velocity}
    \begin{tabular}{c|c|c}
         Model & EHT only & EHT + Millimetron  \\
        \hline
        $5.6 \pm 1.1$ & $-1.0 \pm 30.0$  & $6.2 \pm 3.5$   \\
        \hline
    \end{tabular}
\end{center}
\end{table}

Speaking about M87, time variability for this source is $\sim 220000$~s or $7\times~t_{g}$ -- it is 1000 times higher than for Sgr~A*. Thereby, dynamic imaging observations at HEO perigee are not feasible for this source. Moreover, the shape of the orbit does not allow obtaining for M87 a more or less equivalent $(u, v)$ coverage in the dynamic observation mode. For these reasons, we did not present the results for M87 dynamic imaging at HEO orbit. At the same time, it is worth noting that dynamic imaging of M87 in the perigee can reveal time variation of local events at shorter time scales $t\sim t_g$ near the shadow.

In future works, it is planned to perform the reconstruction of dynamic images by the different methods of dynamic imaging, described in \cite{Johnson2017}. This approach will reduce the impact of the side-lobes in the dirty beam on the resulting reconstructed image.

\section{Conclusions}\label{summ}

The goal of combined Millimetron-EHT space-Earth VLBI simulations was to demonstrate the capabilities of Millimetron, particularly operating in HEO. We have calculated a set of high-elliptical near Earth orbits for Millimetron space observatory, that allow transition from L2 point of Sun-Earth system.

After that, we performed the feasibility study of these orbits using the synthetic space-ground VLBI observations of high resolution imaging of the black hole shadow for M87 and Sgr~ A*.

Not only the results of these simulations show the capabilities of imaging in HEO with Millimetron, but also HEO orbits provide an unique capability of observing Sgr A* in dynamics. It means, that in case the space telescope has a sufficient velocity in the perigee, together with a sufficient number of ground-based telescopes and small baseline projections, one can obtain a rapidly evolving $(u, v)$ coverage. This coverage allows to reconstruct the images with $(u, v)$ sufficient for observing large-scale evolving structures ($\sim 60\times t_{\rm g}$) around the black hole in dynamics.

Results of simulations for space-Earth VLBI imaging of black hole with Millimetron has shown a significant contribution into the spatial and temporal resolution of the VLBI observations with EHT for complex structures of black hole modeled sources. Namely, comparison of the reconstructed images obtained with EHT only and Space-Earth interferometer EHT+Millimetron showed more detailed structure of scattered models (the best examples demonstrated by models 24 and 39) is not resolved for the reconstructed images of the ground only VLBI configuration. Note, that the EHT ground array will likely be extended by the time Millimetron is operational increasing the number of ground and space-ground baseline and improving the $(u, v)$ coverage \citep{Blackburn2019}.

In addition, the resultant angular resolution and $(u, v)$ coverages are evidently improved by Millimetron
as shown in the reconstructed images. Linear angular resolution for the simulated observations of M87 and Sgr~A* for EHT+Millimetron configuration is $\sim 6$ times better, than for the ground only configuration.

Finally the main results of this paper clearly demonstrate the validated possibility of dynamic observations of Sgr~A* with the Space-Earth VLBI baselines. Dynamic observations of black hole vicinity are important for measuring BH physical parameters. Previous capabilities of such observations were discussed only for Space-Space VLBI configurations \citep{Fish2020, Palumbo2019, Roelofs2019}.    

\section*{Data availability}
Data underlying related to the results of simulations in this article will be shared on reasonable request to the corresponding author.

\section*{Acknowledgements}
The authors acknowledge M.~Mo\'scibrodzka for providing the GRMHD models for the described simulations. We would like to thank the reviewer for detailed analysis, valuable comments, remarks and criticism of this paper.
\\
\bibliographystyle{mnras}
\bibliography{m_biblio} 

\bsp	
\label{lastpage}
\end{document}